# Title: Thermal Conductivity of Polymers and Their Nanocomposites

*Xiangfan Xu, Jie Chen*, Jun Zhou*, Baowen Li*

Prof. X. F. Xu, Prof. J. Chen, Prof. J. Zhou

Center for Phononics and Thermal Energy Science, School of Physics Science and Engineering, Tongji University, Shanghai 200092, China
China-EU Joint Lab for Nanophononics, School of Physics Science and Engineering, Tongji University, Shanghai 200092, China
Shanghai Key Laboratory of Special Artificial Microstructure Materials and Technology, School of Physics Science and Engineering, Tongji University, Shanghai 200092, China
E-mail: jie@tongji.edu.cn; zhoujunzhou@tongji.edu.cn

Prof. B. W. Li

Department of Mechanical Engineering, University of Colorado, Boulder, CO 80309, USA

**Abstract**: Polymers are usually considered as thermal insulators and their applications are limited by their low thermal conductivity. However, recent studies showed that certain polymers have surprisingly high thermal conductivity, some of which are comparable to that in poor metals or even silicon. In this review, we outline the experimental achievements and theoretical progress of thermal transport in polymers and their nanocomposites. The open questions and challenges of existing theories are discussed. Special attention is given to the mechanism of thermal transport, the enhancement of thermal conductivity in polymer nanocomposites/fibers, and their potential application as thermal interface materials.

## 1. Introduction

Humanity's insatiable appetite for energy has triggered increasing interest in nanoscale materials and devices that have potential capabilities for improved energy conversion and harvesting. People have been searching for many decades for highly efficient thermoelectric materials that can covert wasted heat into useful electricity. On the other hand, there has been an increasing demand for high thermal conductivity materials that can dissipate waste heat generated by electronic devices during operation. Indeed, the hotspot (accumulated heat) in integrated circuits has become the bottleneck for further development of microelectronics. [1-5] In fact, this hotspot has prevented the operating frequency of microprocessors from going beyond several GHz when their power density exceeds 100 Wcm$^{-2}$. [6, 7]

To remove the heat from the hotspot, one needs to attach the device to a heat sink through a high thermal conductivity material. However, because of surface roughness, only a small portion of the surfaces contact each other when two materials are mated together. Large quantities of gaps and voids filled with air (**Figure 1a**)[8] cause a significant interfacial thermal resistance (ITR) across the contacting surfaces that reduces the heat dissipation efficiency. To reduce the ITR, polymer-based thermal interface materials (TIM) have been used to fill the gaps between the two surfaces. [2, 3]

The total ITR ($R_{TIM}$) consists two parts: $R_{TIM}=R_c+2R_{int}$. The $R_{int}$ is the ITR between the TIM and the substrate materials (e.g., packing materials or heat sink), and $R_c$ is the thermal resistance of the TIM. Considering the thickness and $\kappa$ of commercial TIM, $R_c$ is estimated to be from approximately $10^{-6}$ m$^2$KW$^{-1}$ to $10^{-5}$ m$^2$KW$^{-1}$, which is much larger than the $R_{int}$ (~$10^{-7}$ m$^2$KW$^{-1}$ to ~$10^{-6}$ m$^2$KW$^{-1}$) of organic materials and metal interfaces. [9] This means that $R_c$ dominates the $R_{TIM}$ for current material choices.

However, polymers have a relatively inferior thermal conductivity, ranging from 0.1 Wm$^{-1}$K$^{-1}$ to 0.3 Wm$^{-1}$K$^{-1}$. [10, 11] Various attempts have been exploited to increase the thermal conductivity of polymers (reduce $R_c$) by incorporating high thermal conductivity inorganic materials, such as carbonaceous materials, boron nitride, silicon nitride and metals, to create additional heat pathways.[12-18]

Apart from thermal interface materials, high thermal conductivity polymers are also used for flexible electronics and bioelectronics (**Figure 1b**),[19-21] solar cells (**Figure 1c**),[22-27] heat exchangers,[28] etc.

On the other hand, polymeric thermoelectric materials (**Figure 1d**),[29-31] where lower thermal conductivities are preferred, emerge as new research foci due to their mass-production capabilities and nontoxicity.

Although there is a significant demand for polymers in thermal management applications, the physical picture of thermal transport in polymers and polymer-based nanocomposites is unclear.[11] In fact, a general and systematic theory of thermal transport in bulk polymers is still lacking. The existing theories, such as phonon hopping and minimum thermal conductivity, cannot satisfactorily explain thermal transport in polymers.

Interestingly, although bulk polymers are well known thermal insulators, their thermal conductivity can be enhanced by two to three orders of magnitude due to chain realignment under strain or stretching.[32] For polymer-based nanocomposites, the experimental thermal conductivity results are less than the theoretical calculations due to the unpredictable large ITR between inorganic fillers and the polymer matrix. Moreover, the clustering nature of the particles/wires and their non-uniform distribution give rise to further challenges in modeling the thermal conduction inside the nanocomposites.

In this paper, we give a brief review on recent advances in understanding thermal transport in polymers and their nanocomposites from both experimental and theoretical perspectives. This paper is organized as follows: we first summarize the progress of understanding thermal conductivity in intrinsic polymers in Section II, including theories, molecular dynamics simulations (MD) and recent experimental achievements in enhancing the thermal conductivity of polymer fibers. In Section III, we discuss thermal conductivity in polymer-based nanocomposites; finally, we discuss the challenges and provide an outlook in Section IV.

## 2. Thermal conductivity in polymers
### 2.1. Theories of intrinsic thermal conductivity in polymers

We first focus on bulk polymers without the inclusion of organic fluids and solutions. Thermal transport theory has been established for crystalline solids for over a century, while thermal transport in amorphous and complex materials, such as polymers, remains an open question due to their complicated internal structures. Most polymers are insulators with phonons as the dominant heat carriers. In conducting polymers, charge carriers may also carry heat, however, the validity of the Wiedemann-Franz law is in question, since polarons and bipolarons are major charge carriers, unlike the case in metals.[33-36]

**Figure 2a** and **Figure 2b** show the measured thermal conductivities of typical materials including crystalline silica ($SiO_2$), amorphous glass (a-$SiO_2$), and polymers with different crystallinity, including polyethylene terephthalate (PET), polyethylene (PE) and amorphous polymer polymethyl methacrylate (PMMA). The temperature-dependent thermal conductivity of inorganic crystals has been well explained by the increase of the specific heat at low temperatures ($\kappa \propto T^3$) and the Umklapp phonon scattering process at high temperatures ($\kappa \propto T^{-1}$). However, the case of amorphous materials, both inorganic and organic, is different. The thermal conductivity first increases with $T^2$, then undergoes a plateau transition to increase monotonically, and finally saturates. The temperature-dependent behaviors of semi-crystalline polymers are sensitive to the crystallinity, which is noted by X. The plateau diminishes with the increasing X value (**Figure 2b**), while the monotonically increasing trend disappears when X is large, as shown in **Figure 2a**.

Amorphous polymers possess low thermal conductivity due to the lack of crystalline structures. Defects (such as voids, entanglements, chain ends and impurities) give rise to excess scattering events and reduce the thermal conductivity.[32] Simplified models based on relaxation times and Debye approximations are still widely used. The thermal conductivity is then given by

$$\kappa(T) = \frac{1}{3}\sum_j \int C_j(\omega) v_j \ell_j(\omega) d\omega, \tag{1}$$

where $C_j(\omega)d\omega$ is the specific heat contribution of phonons with branch index j and frequency $\omega$, $v_j$ is the group velocity of the phonons and $\ell_j(\omega)$ is the mean free path (MFP). Different scattering mechanisms are revealed from the form of the MFP and result in different temperature dependent behaviors of $\kappa$. At very low temperatures, resonant scattering by two-level states is dominant and the phonon MFP is approximately proportional to $\omega^{-1}$. Therefore, the thermal conductivity follows $\kappa \propto T^2$.[10] The plateau occurring at several Kelvin is attributed to the delocalization-localization transition, indicating that $\kappa$ saturates from excited delocalized phonons.[37] Above the plateau, $\kappa$ continues to increase and saturates at above ~100K. The "minimum thermal conductivity model" articulates that the thermal conductivity approaches the "amorphous limit", which can be derived from **Equation (1)** assuming the MFP is reduced to half the wavelength.[38] However, very little is known about the microscopic origin of the two-level states and localized modes in polymers. The "minimum thermal conductivity model" is also challenged by the rationality of the Debye approximation in polymers with strong anharmonicity and its failure to recover the temperature dependence at low temperatures.

**2.2 MD simulation of thermal conductivity in polymers**

The MD simulation is a powerful modeling technique that can handle complex systems with atomic level information based on classical mechanics. It helps to provide a molecular level understanding of the underlying mechanisms, serving as an important aid to experiments to study the structure-property relationship in various materials. The MD simulation only requires the material structure and suitable interatomic potentials as the input, avoiding assumptions that are often used in the theoretical models. Many force fields for polymers exist in the literature, such as the polymer consistent force field (PCFF)[39, 40], the condensed-phase optimized molecular potentials for atomistic simulation studies (COMPASS)[41], and the adaptive intermolecular reactive bond order (AIREBO) potential.[42]

There are two types of MD simulations widely used to calculate the thermal conductivity of polymers. One is the non-equilibrium MD (NEMD) simulation, in which a temperature difference is introduced into the simulation domain and the thermal conductivity is computed according to Fourier's law as $\kappa=-J/\nabla T$, where $J$ and $\nabla T$ are heat flux and temperature gradient across the system, respectively. The heat flux is defined as the energy transported per unit time across the unit area. The non-equilibrium state can be established either by applying two thermostats [43] at different temperatures to maintain a constant temperature at the two ends of the system, or by artificially swapping atom velocities in different regions to impose a constant heat flux [44, 45], also known as the reverse non-equilibrium MD (RNEMD) method.

Another approach to compute the thermal conductivity is based on the Green-Kubo formulism in equilibrium MD (EMD) simulations as

$$\kappa_{\alpha\beta} = \frac{V}{k_B T^2} \int_0^\infty \langle J_\alpha(t) J_\beta(0) \rangle dt, \qquad (2)$$

where $k_B$ is the Boltzmann constant, $V$ and $T$ denote the volume and temperature of the system, respectively, $J_\alpha$ is the heat flux in the $\alpha$ direction, and the angular brackets denote the ensemble average. The heat flux vector can be written as

$$\mathbf{J}(t) = \frac{1}{V} \frac{d}{dt} \sum_{i=1}^{N} \mathbf{r}_i E_i, \qquad (3)$$

where $\mathbf{r}_i$ and $E_i$ are the position and total energy of the $i$-th atom, respectively.

One peculiar discovery through MD simulations is the divergent thermal conductivity in the polymer chain. Bulk amorphous polymers are often used as thermal insulators [46]. Contrary to this common wisdom, Henry and Chen [47, 48] discovered through MD simulations that the thermal conductivity of polymers is not intrinsically low. Their EMD simulation results showed that the thermal conductivity of individual polyethylene chains can exceed 100 Wm$^{-1}$K$^{-1}$ if the chain is longer than 40 nm, and even show a divergent thermal conductivity with the increasing length in certain simulations.[47] This finding was later confirmed by experimental measurements of polyethylene nanofibers.[32] Each phonon mode is specified by a wavevector *k* and polarization *p*. In contrast to three-dimensional systems where the cross correlations (*k*≠*k'*, or *p*≠*p'*) between different phonon modes are negligible, they further found that in one-dimensional systems, such as individual polyethylene chains, the cross correlation between the longitudinal-acoustic (LA) phonons is significant and does not fully attenuate.[48] By expressing the heat flux in terms of the phonon modes and plugging it into the Green-Kubo formula, the thermal conductivity can be written as [48]

$$\kappa = \frac{1}{V}\sum_k \sum_p \sum_{k'} \sum_{p'} vv' \sqrt{CC'} \int_0^\infty \frac{\langle \delta n(t) \delta n'(t+t') \rangle}{\sqrt{\langle \delta n^2(t) \rangle \langle \delta n'^2(t) \rangle}} dt', \qquad (4)$$

where *v* is the phonon group velocity, *C* is the specific heat, and *δn* is the deviation from the average phonon occupation number. When the polymer chain is under deformation, both the sound velocity and heat capacity will be affected.[49, 50] Their thorough analysis [48] revealed that the persistent cross correlation between the LA phonons, which is the largest contribution to the thermal conductivity, substantially contribute to the thermal conductivity as the cross terms in **Equation (4)**, leading to the divergent thermal conductivity in single polyethylene chains. Liu and Yang [51] further found that the thermal conductivity of single extended polymer chains increases with its length as the power law $\kappa \sim L^\gamma$, and the exponent *γ* depends strongly on the monomer type.

Based on MD simulations, various aspects of the thermal transport in polymers and their nanocomposites have been investigated. By examining different polymer nanofibers, Zhang *et al.*[52] found that the *π*-conjugated nature is important to simultaneously achieve high thermal conductivity and good thermal stability in polymers, due to the suppressed segmental rotations and large phonon group velocity. For bulk amorphous polymers, a larger density and extended chain morphology are found to have a higher thermal conductivity, mainly caused by the covalent bonding contribution.[53] In contrast, the virial contribution from van der Waals interactions is found to dominate the thermal conductivity in the crosslinked polymer network.[54] For randomly coiled polymer chains, Liu and Yang [55] found through MD simulations that the polymer chains gradually align themselves along the stretching direction, manifested by the increase of the orientational order parameter with the applied strain. Since the well-aligned polymer chain has an intrinsically high thermal conductivity, an improved chain alignment from the strain increases the thermal conductivity along the stretching direction, which is larger when the polymer is stretched more slowly. Grafting and covalent functionalization are found to be effective ways to improve the interfacial heat conduction across polymer/inorganic solid interfaces[56-59], which usually increases with the grafting density. Hu *et al.*[60, 61] found that neither the acoustic mismatch nor diffusive mismatch model are capable of quantitatively predicting the scattering coefficient at the silicon/amorphous polyethylene interface, and the interfacial bonding strength has a more pronounced impact than the solid modulus in reducing the ITR.

The phonon gas picture, widely used to describe lattice vibrations in crystals where atoms are arranged in an ordered and periodic fashion over a large length scale, has exhibited remarkable success in modeling and interpreting thermal transport in various crystalline solids. In this picture, phonons are plane wave modulated propagating modes with well-defined phase and group velocities, and there exists a clear definition of the phonon dispersion relation in the crystals. However, in amorphous/disordered materials, such as bulk polymers, where the long-range periodicity or compositional homogeneity is lacking, these phonon-related properties, such as the group velocity and relaxation time, are not well defined. This brings into question the use of the phonon gas picture to model the thermal transport in amorphous solids.

The seminal works by Allen and Feldman[62, 63] (AF) proposed an alternative harmonic model to describe the heat transport in amorphous solids as

$$\kappa = \frac{1}{V} \sum_i C_i(T) D_i, \tag{5}$$

where $C_i(T)$ is the specific heat of the $i$th mode and $D_i$ is the temperature-independent mode diffusivity. In this model, heat is carried by the off-diagonal elements of the heat current operator, which has a nonzero contribution due to the nonzero off-diagonal density matrix induced by the temperature gradient. Allen *et al.*[64] introduced new mode classifications in non-periodic systems, namely, propagons (ballistically propagating phonons), diffusons (diffusive phonons), and locons (localized phonons). These modes can be distinguished by calculating the eigenvector periodicity.[65] In amorphous silicon, they found that only 3% of modes are locons, while diffusons fill 93% of the spectrum, and the remaining 4% of modes are propagons.[64] Their calculation results revealed that the re-increase in the thermal conductivity of amorphous silicon above the plateau region is attributed to the contribution from diffusons.

There have been growing efforts in recent years to study the thermal properties of amorphous solids.[66-68] Zhu and Ertekin[69] presented a generalized model that describes the vibrational transport in low-dimensional and disordered materials, which is a generalization of the Debye-Peierls and Allen-Feldman schemes. Based on the AF model, Kommandur and Yee [70] proposed an empirical model that can well predict the temperature dependent thermal conductivity of amorphous polymers, especially for the plateau-like transition at temperatures approximately 10 K. The plateau-like transition at low temperatures is not well accounted for using existing models, such as the phonon gas picture. Lv and Henry[71] derived a new method termed the Green-Kubo modal analysis (GKMA) as

$$\boldsymbol{Q}(t) = \sum_n^{3N} \boldsymbol{Q}(n,t) = \sum_n^{3N} \frac{1}{V} \sum_i \left[ E_i \dot{x}_i(n,t) + \sum_j \left( -\nabla_{r_i} \Phi_j \cdot \dot{x}_i(n,t) \right) r_{ij} \right], \qquad (6)$$

$$\kappa = \frac{V}{k_B T^2} \sum_{n,n'} \int_0^\infty \langle \boldsymbol{Q}(n,t) \cdot \boldsymbol{Q}(n',0) \rangle dt, \qquad (7)$$

where $E_i$ and $\Phi_i$ are, respectively, the total energy and the potential energy of the $i$th atom, and $\dot{\mathbf{x}}$ is the displacement vector. The predicted temperature dependent thermal conductivity for amorphous silicon using the GKMA method shows good agreement with the experimental results.[71] Different from the AF model, which is the harmonic approximation, the GKMA method includes the anharmonic effect, and can treat the ordered crystals, disordered materials, and interfaces.[71-73]

**2.3 Enhanced thermal conductivity in polymer fibers**

The low thermal conductivity of bulk polymers is due to the intrinsic mechanisms, such as weak chain bonds, chain twisting, entanglement, chain ends, and extrinsic multiple scatterings such as voids, defects, etc. (**Figure 2c** and **2d**).[52, 74] The thermal conductivity increases with the sound velocity and elastic modulus, which are related to the chain bond. To enhance the thermal conductivity in polymers, it is straightforward to change the chain bond to modify the sound velocity and elastic modulus.[38, 75-77] A recent study[38] (**Figure 2e**) shows that polymers with weak intermolecular interactions, such as polystyrene (PS) and PMMA, have the lowest thermal conductivity of 0.1 $Wm^{-1}K^{-1}$ to 0.2 $Wm^{-1}K^{-1}$, followed by three hydrogen-bonded polymers (~0.3 $Wm^{-1}K^{-1}$ to 0.5 $Wm^{-1}K^{-1}$). The thermal conductivity of ion-bonded polymers has the highest value of ~0.5 $Wm^{-1}K^{-1}$ to 0.7 $Wm^{-1}K^{-1}$. Nevertheless, all the measured bulk polymers possess a thermal conductivity that is much lower than that in amorphous silicon dioxide. Interestingly and contradiction to conventional wisdom, the thermal conductivity of crystalline PE has a high value of approximately 20 $Wm^{-1}K^{-1}$ [75], indicating that the relatively low thermal conductivity in polymers is related to the extrinsic scatterings mentioned above.

The re-alignment of polymer crystallites is proposed to enhance the mechanical strength, which increases the thermal conductivity (**Figures 2f** and **2g**).[78-84] Choy *et al*. [78, 80, 85] demonstrated that the thermal conductivity of fiber-reinforced plastics (FRPs) consisting of Spectra 100 and PE have a room temperature thermal conductivity of ~30 $Wm^{-1}K^{-1}$. The work carried out by Fujishiro *et al*. demonstrates that both micro-fiber bundles and FRPs have a thermal conductivity of approximately 60 Wm$^{-1}$K$^{-1}$ at 260 K for both Dyneema and Zylon.[86] Both groups observed a two order of magnitude enhancement in the thermal conductivity of polymers fibers with diameters limited to the micro-scale. An astonishing work presented by Shen *et al*.[32] recently found that the thermal conductivity of PE fibers with diameters of 131±12 nm can be as high as 104 Wm$^{-1}$K$^{-1}$ (compared to ~0.35 Wm$^{-1}$K$^{-1}$ in bulk PE). This is understandable because the Green-Kubo approach based MD simulations shown that the thermal conductivity in a single PE chain approaches 350 Wm$^{-1}$K$^{-1}$.[51, 55]

An experiment carried out on electron spinning [87, 88] Nylon-11 nanofibers demonstrates that the thermal conductivity starts to increase when the nanofiber diameter is smaller than 200 nm, with the highest value of ~1.6 Wm$^{-1}$K$^{-1}$ in 73 nm diameter nanofibers. This diameter-dependence behaves quite different from inorganic nanowires where the thermal conductivity reduces with the decreasing diameter due to surface scatterings. This indicates that the enhancement observed is due to structural changes, which was confirmed by high-resolution wide-angle X-ray scattering.[88] In terms of the relatively lower thermal conductivity observed in Nylon-11 compared with that in the PE nanofibers, the degree of crystallinity of Nylon-11 is only 35%-40% compared to that of 80%-90% in PE, apart from the fact that the intrinsic Young's modulus in Nylon (~25 GPa) is one order of magnitude smaller than that in PE (~240 GPa).[88] Semi-crystalline polymers consist of crystalline units and amorphous parts (**Figure 2f**). It is not surprising that a semi-crystal can be treated as a two-phase material and explained well by the effective medium model considering both the crystalline orientation and anisotropic thermal transport through intra-chain bonding and inter-chain interactions.[10] This model coincides with the observed enhancement of the thermal conductivity with the increase in the crystallinity or orientation after drawing or spinning.[10]

The amorphous polymer nanofibers follow the same behavior as that in crystalline or semi-crystalline polymers. Singh *et al*. showed that pure polythiophene nanofibers have a thermal conductivity up to ~4.4 Wm$^{-1}$K$^{-1}$ while remaining amorphous.[74] This enhancement in thermal conductivity enables vertically aligned arrays of polythiophene to be used as TIM with a total thermal resistance as low as 12.8 ± 1.3 mm$^2$KW$^{-1}$.[74]

It is worth noting that the mass production of aligned polymer nanofiber bundles is highly desired for scale-up applications where anisotropic thermal conduction is needed. Various attempts, e.g., static-electron spinning[87], AAO templates[74, 89] and small molecule epitaxy,[90] have been used to fabricate high orientation polymer fibers or films, although further efforts are needed to exploit their applications as heat spreaders or heat dissipaters.

Apart from artificial polymers mentioned above, the natural organismal fibers, such as silkworm and spider silk, also possess high thermal conductivity.[91] The *NC* dragline spider silk has a superior high thermal conductivity of ~ 349 Wm$^{-1}$K$^{-1}$, which is greater than silicon and most metals. This value can be further increased up to ~ 416 Wm$^{-1}$K$^{-1}$ under ~20% strain due to the intra-chain and inter-chain H-bond breaking and improved chain alignment. The strong nonlinear strain-dependent phononic bandgap and phonon dispersion [92] could be responsible for the strain-induced thermal conductivity.

## 3 Thermal conductivity in polymer nanocomposites

### 3.1 Theories of thermal transport in polymer nanocomposites

Compositing is a promising method to modify the thermal conductivity of organic materials by the incorporation of fillers with a high thermal conductivity.[93-95] Electrically insulating fillers such as boron nitride, boron carbine, and aluminum oxide are commonly used when insulator nanocomposites are required in certain applications. Electrically conductive fillers such as carbon nanotubes (CNT), graphene, metal nanowires and metal nanoparticles are used for applications without an insulation requirement. Heat will flow through both the organic matrix and the inorganic fillers. As a result, the effective thermal conductivity of such composites is a function of the intrinsic thermal conductivity of the constituents, loading fraction, filler shape and size, filler dispersion and ITR. These parameters make it difficult to predict the thermal transport properties in the composites.

Until recently, many theoretical models based on the effective medium theory (EMT) have been proposed to estimate the thermal conductivity of composites. The EMT treats the composites as a uniform medium, and was first proposed by Maxwell Garnett (MG), and then improved by Bruggeman; they are referred to as MG EMT and Bruggeman EMT, respectively.[96, 97] Both of these approaches neglect the ITR. The former is valid for small loading fractions, while the latter is applicable for all loading fractions. However, most of the measured thermal conductivities are much less than the theoretical predictions, which is attributed to large ITR between the fillers and the organic matrix, and the contact thermal resistance between the connected inorganic fillers.[98, 99] For example, in the case of the polymer-based nanocomposites embedded with randomly dispersed CNTs or graphite, a point contact with a small contact area will be formed between two cross CNTs or rigid graphite nanoplatelets. The weak interaction between the contacted fillers could lead to a high contact thermal resistance.[100]

Many attempts have attempted to consider the ITR and fillers with different geometries and topologies, such as aligned continues fibers, spheres and completely misoriented ellipsoidal particles.[101, 102] These models provide a better agreement with experiments, which indicates the importance of the interface effect in determining the effective thermal conductivity of the composite materials. To further understand the physical mechanism of thermal transport in the composites, the acoustic mismatch model (AMM) and the diffuse mismatch model (DMM) were first proposed to estimate the ITR between the liquid-solid and solid-solid interface.[103] However, both the AMM and DMM could not accurately describe the phonon interfacial scattering process. Some researchers elucidated the interfacial thermal resistance between organic and inorganic materials by performing MD simulations[60], and found that the bonding strength across the interface plays an important role in the resistance. Recently, Lu *et al.*[104] studied the ITR in a metal-insulator interface by considering the electron-phonon coupling, and indicated that the electron-phonon coupling at the interface is also a significant channel for thermal transport. Accurately determining the ITR at the organic/inorganic interface will be useful to understand the thermal transport mechanism in such composites. These topics deserve further investigations.

In addition to EMT, it is also claimed that the thermal percolation behavior can be observed in organic/inorganic composites at the electrical percolation threshold.[105] When the filler loading fraction is low, the fillers are discretely dispersed or form finite clusters in the polymer matrix. However, when the loading fraction is above the electrical percolation threshold, isolated fillers will contact each other and form an infinite cluster. In this case, the thermal conductivity is largely governed by the conductive network constructed by the fillers. However, the thermal properties are very different from the electrical properties in composites. The electrical properties of the organic matrix are so poor that they can be neglected, and only the conductive fillers contribute to the electrical conductivity.[106] Whereas in the thermal case, there is no absolute thermal insulator, so the contribution from the thermal transport through the organic matrix cannot be ignored. More fundamental reasons for this experimentally observed thermal conductivity behavior are attributed to the role of the ITR or the intrinsic thermal conductivity of the constituents, which is related to the geometric factors, such as the size and porosity of the applied fillers.[107] There is still no clear evidence for thermal percolation behavior. Therefore, it remains an issue under intense debate whether the enhancement in such composites should be interpreted in light of thermal percolation or EMT.

## 3.2 Experimental progress

Here, we focus on recent advances in graphene and CNTs-based nanocomposites. For achievement of other fillers, please refer to references [108-117] for more details.

Graphene, graphite and CNTs have outstanding mechanical and thermal properties. The measured thermal conductivity of a suspended single layer of graphene is around ~5300 $Wm^{-1}K^{-1}$; the value decreases with the increasing number of layers and approaches that of bulk graphite (~1800 $Wm^{-1}K^{-1}$ to 2100 $Wm^{-1}K^{-1}$) very quickly.[118-123] However, the thermal conductivity of these carbon-based nanocomposites is much smaller than the expected value. This is likely caused by but no limited to the following reasons: difficultly in increasing the filler volume fraction, size-dependent thermal conductivity of the fillers, and significant weak coupling induced thermal contact resistance between the fillers and their interfaces with the matrix (polymers).

**Figure 3a** shows the thermal conductivity of graphene-based polymer (epoxy resin) nanocomposites as a function of the volume fraction. The highest value obtained was up to 12.4 $Wm^{-1}K^{-1}$ with the graphene volume fraction reached 25%.[124] The thermal conductivity increases slowly with small volume fractions, followed by a sharp turning when the volume fraction is between 15% and 20%.[124] This percolation-like[125] behavior is related to the additional pathway between the connecting graphene nanoplatelets when the volume fraction increases from 13% to 19% (**Figure 3b**).[124, 126] When the thermally conductive fillers randomly disperse in the low thermal conductivity matrix, a percolation network is likely to be established when the filler volume fraction is sufficiently large, similar to the percolation in electrical networks.

In addition to the volume fraction, [127, 128] the thermal contact resistance between fillers and their interfaces with the matrix is response for the low thermal conductivity.[129] Surface functionalization provides an effective way to reduce the thermal contact resistance by increasing the coupling between the CNTs and the polymers. However the functionalization induces secondary damage to the CNTs, which decreases their intrinsic thermal conductivity.[93] Alternatively, Cui *et al.* fabricated a silica shell onto the surface of CNTs to increase the interface coupling between CNTs and polymers without affecting the intrinsic thermal properties of the fillers.[130] In addition, increasing the contact area between fillers and polymers through modifying the aspect ratio of the filler also provides an effective method to reduce the thermal contact resistance.[131] **Figure 3c** shows the thermal conductivity of nanocomposites with fillers of different aspect ratios and dimensions, including carbon black (CB), graphitic microparticles (GMP), single-walled CNTs (SWNTs) and graphite nanoplatelets (GNP). The four-layer GNP provide the highest thermal conductivity up to 6.44 $Wm^{-1}K^{-1}$ (loading of 25%), which is 30 times larger than that of the matrix polymers and surpasses the performance of conventional fillers of 70 vol% loading.[131] This outstanding thermal performance is attributed to the high aspect ratio, two-dimensionality, etc.

## 4 Outlook

This review presents recent achievements of thermal conductivity in polymers and polymers-based nanocomposites from both experimental and theoretical perspectives, aiming at hunting for superior thermal conductivity polymers as potential candidates for thermal management. It is now understood that the low thermal conductivity of polymers originates from the intrinsic weak chain bonds and extrinsic multiple scattering sources, although the thermal conductivity in such nanocomposites depends on various physical parameters, such as the filler shape and size, filler thermal conductivity, loading volume fraction, ITR of the fillers and their interfaces with the polymers, etc. However, a comprehensive quantitative theory to describe the thermal properties is still lacking. We conclude this review by pointing out several important issues and challenges that deserve further investigation:

a) The intrinsic thermal conductivity of polymers provides a baseline in determining the thermal conductivity of the related nanocomposites. The enhancement of the intrinsic thermal conductivity through crystallization or chain alignment will provide a deeper understanding of thermal properties in polymers. However, the existing theories, including the minimum thermal conductivity[132-134] theory and hopping mechanism[135, 136], fail to quantitatively describe this enhancement.

b). The non-negligible ITR between the fillers and their polymer interfaces is a bottleneck for further increasing the thermal conductivity in polymer nanocomposites. Variable synthesis methods are needed to enhance the interface contact by modifying the weak coupling from van der Waals bonds to covalent bonds, or even hydrogen bonds and ionic bonds while maintaining the intrinsic thermal conductivity of the fillers.

c) Substrate/organic contamination reduces the thermal conductivity in carbon materials.[137-139] It is well known that the substrate or organic residues on surfaces could reduce the thermal conductivity of graphene to below 10 $Wm^{-1}K^{-1}$.[140] This is probably the main reason why carbon-based nanocomposites have much smaller thermal conductivities than that of metal nanowire-based nanocomposites under the same volume fractions, although the thermal conductivity of graphene [118, 141] and CNTs [142] are believed to be much higher than all of the metals.[143]

d) New theories are expected. The current EMT can only explain the experiments with limited loading volume fraction, whereas the ITR between fillers and their interfaces with polymers dominate the thermal conduction. With the volume fraction further increasing, new heat pathways crossing the filler interfaces will be formed. It is very interesting to understand the effect of the percolated network on the thermal conductivity of nanocomposites.

Last but not lease, we shall mention that the ultra-low thermal conductivity is essential to obtain a large figure of merit to achieve high performance thermoelectric materials,[31, 36,144-145]. A deeper understanding of the thermal transport in polymers will shed light on designing high efficient thermoelectric materials.

**Acknowledgements**

This work was supported by the National Key R&D Program of China (No. 2017YFB0406000) and by the Shanghai Committee of Science and Technology in China (No. 17142202100 & No. 17ZR1447900 & No. 17ZR1448000)

[145] X. Su, P. Wei, H. Li, W. Liu, Y. Yan, P. Li, C. Su, C. Xie, W. Zhao, P. Zhai, Q. Zhang, X. Tang, C. Uher, *Adv. Mater.* **2017**, *29*, 1602013.

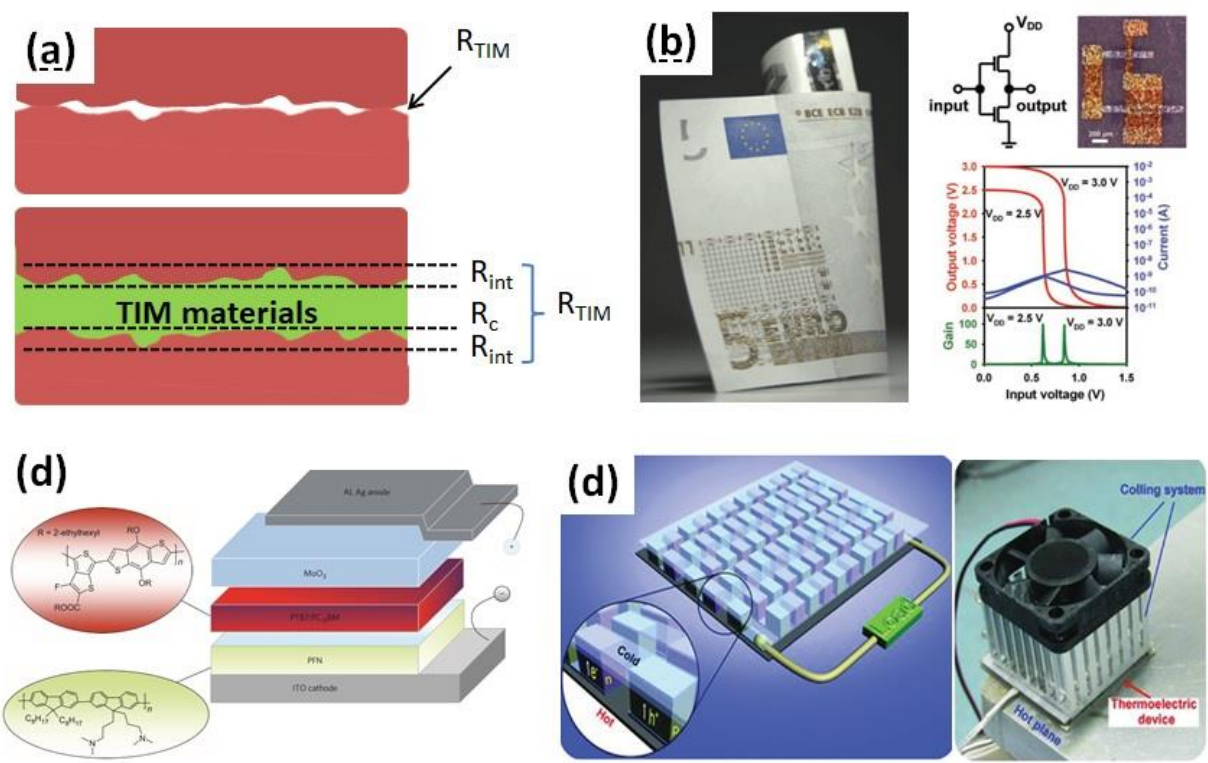

**Figure 1.** (a) Thermal interfacial material fills the voids and gaps to reduce the interfacial thermal resistance. (b) Photograph of a 5-Euro note with arrays of organic transistors and circuits. Reproduced with permission.[19] Copyright 2011 published by WILEY-VCH. (c) Schematic of the inverted type polymer photovoltaic cells. Reproduced with permission.[27] Copyright 2012 Published by Springer Nature. (d) Organic thermoelectric materials. Reproduced with permission.[30] Copyright 2012 published by WILEY-VCH.

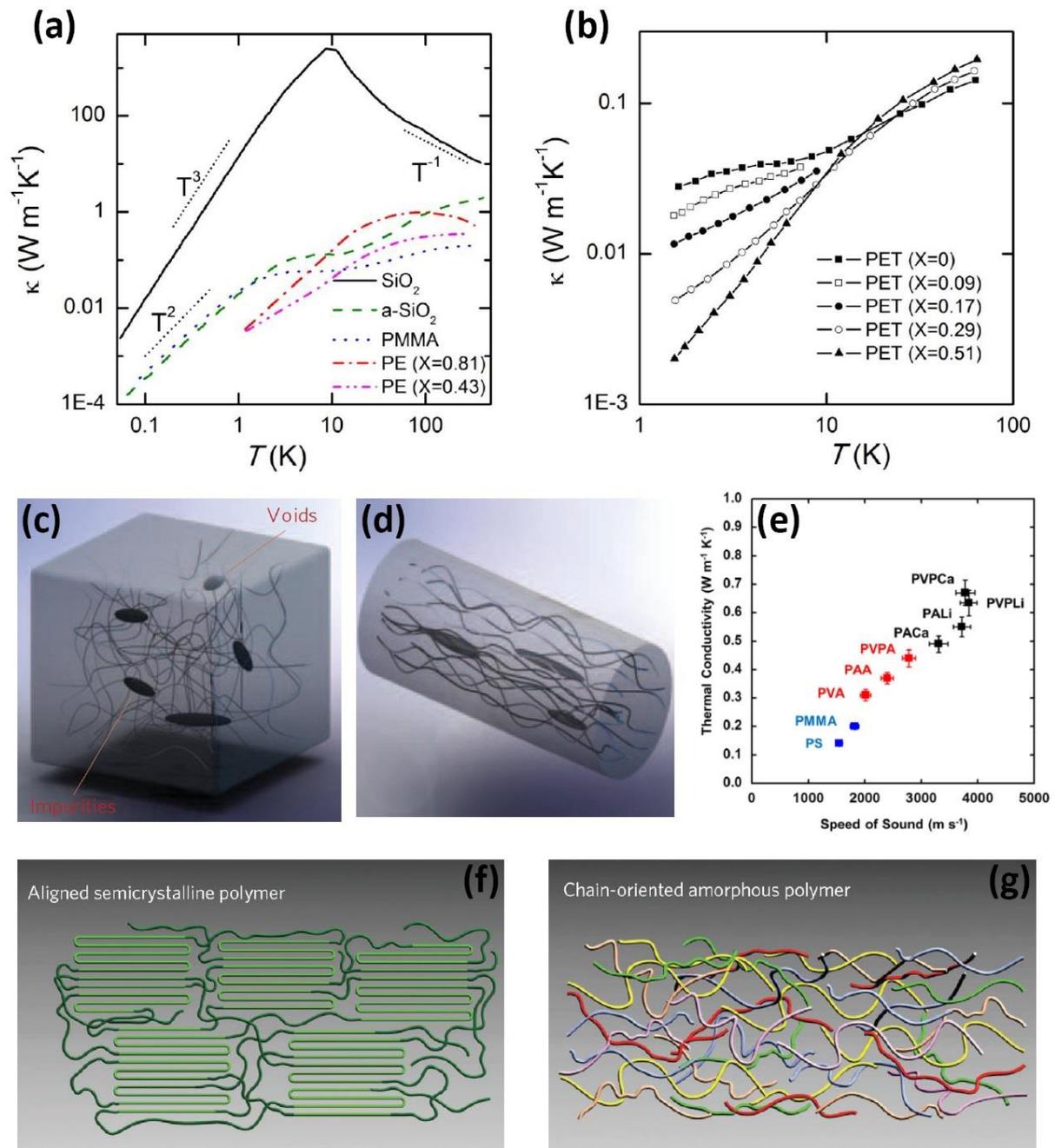

**Figure 2** (a) Thermal conductivity of crystalline and amorphous SiO$_2$, poly(methyl methacrylate) (PMMA) and polyethylene (PE) with different crystallinities X. Dotted lines indicate the power law dependence. (b) Thermal conductivity of poly(ethylene terephthalate) (PET) with different crystallinities X. Reproduced with permission.[10] Copyright 1977 Published by Elsevier Ltd. (c) Extrinsic multiple scattering sources, such as voids, defects, chain twisting, entanglement, and chain ends, in bulk polymers. (d) Stretched polymer fiber. Reproduced with permission.[32] Copyright 2010 Published by Springer Nature. (e) Thermal conductivity of polymers with different interchain bonds. Reproduced with permission.[38] Copyright 2017 Published by American Physical Society. (f) and (g) Microstructures of the chain-orientation in semi-crystalline and amorphous polymers. Reproduced with permission.[74] Copyright 2014 Published by Springer Nature.

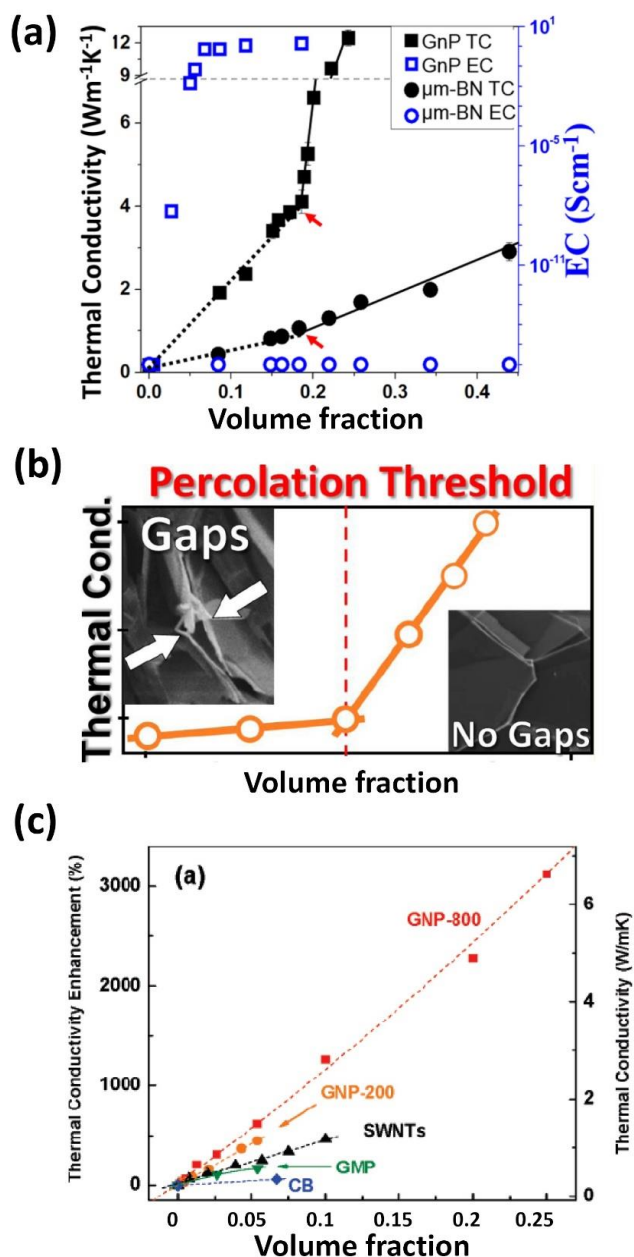

**Figure 3** (a) Thermal conductivity and electrical conductivity of epoxy resin based nanocomposites as a function of the loading volume fraction of graphite nanoplatelets (GnP) and micron size boron nitride (BN). Reproduced with permission.[124] Copyright 2015 Published by the American Chemical Society. (b) Percolation-like behavior with the loading fraction increased from 13% to 19%, during which the pathway between two connected fillers increases due to the extinct gas. Reproduced with permission.[126] Copyright 2015 Published by the American Chemical Society. (c) Thermal conductivity enhancement of the epoxy with different fillers: carbon black (CB), graphite microparticles (GMP), single-walled carbon nanotubes (SWNTs) and graphene nanoplatelets (GNP). Reproduced with permission.[131] Copyright 2007 Published by the American Chemical Society.